\documentstyle[twocolumn,pra,aps,epsf,epsfig]{revtex}

\newcommand{\be}{\begin{equation}}
\newcommand{\ee}{\end{equation}}
\newcommand{\bea}{\begin{eqnarray}}
\newcommand{\eea}{\end{eqnarray}}

\newcommand{\Mat}[1]{\widehat{#1}}

\title{Free Energy Self-Averaging in Protein-Sized Random
Heteropolymers}
\author{Jeffrey Chuang$^1$, Alexander Yu. Grosberg$^{2,3}$, and Mehran
Kardar$^{1,4}$}

\address{
$^1$Department of Physics, Massachusetts Institute of Technology,
Cambridge,
Massachusetts 02139 \\
$^2$Department of Physics, University of Minnesota, Minneapolis,
Minnesota 55455 \\ $^3$Institute for Biochemical Physics, Russian
Academy of Sciences, Moscow 177234, Russia \\ $^4$Theoretical
Physics, University of California,
Santa Barbara, California 93106\\
}

\address{{\em \bigskip
\begin{quote}
Current theories of heteropolymers are inherently {\it
macro}scopic, but are applied to folding
proteins which are only {\it meso}scopic.  In these theories,
one computes the averaged free energy over sequences, always
assuming that it is
self-averaging -- a property well-established only if a system
with quenched disorder is macroscopic.
By enumerating the states and energies of compact 18,
27, and 36mers on a simplified lattice model with an ensemble
of random sequences, we test the validity of the
self-averaging approximation. We find that fluctuations in the free
energy between sequences are weak, and that
self-averaging is a valid approximation at the length scale of real
proteins.  These results validate certain
sequence design methods which can exponentially speed up
computational design and greatly simplify experimental realizations.
\end{quote}
}}

\begin{document}
\maketitle

Protein folding remains one of the most challenging problems in polymer
physics \cite{RMP,Shakhnovich,Wolynes,Dill_Review}. The phenomenon is
straightforward -- at low temperature
a heteropolymer chain freezes into a single configuration. However,
the relationship between a chain's monomer sequence and the
thermodynamics
of its transition is complex. As a result, current
theories of heteropolymer freezing resort to certain assumptions which
have not been adequately tested, one of the most basic being
self-averaging of the free energy.

Self-averaging is a property of many disordered
systems, stating that the free energy of a system of size $N$ with
quenched disorder is independent of the particular realization of the
disorder,
to within variations of order of $\sqrt{N},$ which are relatively
negligible as $N\rightarrow \infty$.
This property can be rigorously proved for a broad range of models in
macroscopic disordered systems\cite{Gredeskul}.
Fundamentally, it stems from the
independence of sub-regions in the $N \rightarrow \infty$ thermodynamic
limit.
In the context of heteropolymers, it states that a random
heteropolymer's
free energy is independent of its sequence, i.e.
\be
F(seq,T) \simeq \left\langle F(seq,T)\right\rangle_{seq},
\ee
where $\left\langle \cdots\right\rangle_{seq}$ indicates an average 
over sequences.
There have been some proofs of self-averaging for certain heteropolymer
models in the
 $N \rightarrow \infty$ limit \cite{Orlandini}. Proteins, however, are
mesoscopic objects, and it is unclear whether self-averaging
applies at the lengths of $N$ not more than several hundred
monomers found in proteins.

Self-averaging in heteropolymers is important for two main
reasons. {\em First}, it is relevant to the theoretical
understanding of protein folding. Starting from
\cite{Shakhnovich,Wolynes}, key modern theories of heteropolymers,
reviewed recently in \cite{RMP}, compute the averaged free energy
of the system, implicitly neglecting sequence-dependent variations
in the manner of Eq.~(1). In particular, self-averaging is an
element of the replica method, and is used in the
derivation of the Random Energy Model for heteropolymers
\cite{Shakhnovich}. {\em Second}, self-averaging is an important
assumption of certain sequence design methods such as
``imprinting'' \cite{Vijay_Imprint} and ``sequence
selection\cite{Selection},'' ideas from which have been used for
de novo protein and ligand design \cite{Hecht}. These methods 
have proven useful both experimentally \cite{EnokiPRL} and
computationally. Unfortunately, computational design methods that
do not assume self-averaging\cite{Dima} require vastly more
calculation time.
To design a heteropolymer sequence to fold into a certain
conformation $*$ at temperature $T$, one should minimize the
quantity $E(seq,*)-F(seq,T)$ over all sequences. If self-averaging
is not assumed, one must calculate the energy of {\em all}
conformations for each sequence tested to determine $F(seq,T)$.
However, if self-averaging is valid, then the $F$ term can be
ignored and design can be carried out by evaluating the energy of
each sequence in just the one conformation $*$. This exponentially
speeds up the design procedure.

{\em In vitro} experiments have not yet provided sufficient
evidence to verify self-averaging in random peptide chains.
In the experiments that have studied random amino acid sequences,
there have not been any obvious trends in the behavior
\cite{Sauer_Cooperative}, due to the difficulty of
such experiments and consequent lack of data.
However, using a computer simulation, we are able
to sample many more sequences than can be analyzed feasibly {\em in
vitro},
and thus determine whether the property
of free energy self-averaging over sequences is valid for
heteropolymers.
We perform a scaling comparison of the exact free energy and
other parameters for several three-dimensional lattice heteropolymers
of different size. We then extrapolate our data to determine
the validity of free energy self-averaging for protein-sized polymers.

In order to study the thermodynamics of random heteropolymers, we
perform an exact lattice enumeration of the states of compact
polymer chains  of several different lengths. For each length, we
examine many random sequences made up of two monomer species,
$\alpha$ and $\beta$. All included sequences have the same number
of monomers of type $\alpha$ and type $\beta$, so as to remove any
concentration dependence. For each sequence, we evaluate its
energy in all possible compact conformations. Using this
information, we then calculate the free energy $F(T)$ of each
sequence. To determine whether self-averaging is valid, we first
compare the average and the standard deviation of the free energy
over the examined sequences, and then examine the dependence of
these quantities on the chain length $N$.
Our method of enumeration is in contrast with  other works
which use Monte Carlo sampling of states to determine averaged
thermodynamic properties \cite{RMP,Bastolla}.
By doing a full enumeration, we
are able to separate thermodynamic properties of the system
from kinetic effects.
In practice, the method is similar to the procedure used in
studies of designability \cite{LiTangWingreen}, although we
only examine a finite sample of sequences, rather
than testing every possible one. In our case, the focus is 
the complete free energy versus temperature curve, rather 
than just the ground state conformation for each sequence.

We use a standard model in which monomers are
placed at lattice positions ${r_i}$, and subject to an energy \be
E = \sum_{i < j }^N B_{s_i s_j} \Delta(r_i - r_j), \ee where $i$
and $j$ run over the monomers in the chain, and $s_i$ indicates
one of the species ($\alpha$ or $\beta$) of the monomer $i$ for a
particular sequence $\{s_i\}$. Contact interactions are enforced
by setting $\Delta(r_i - r_j) = 1$, if the $r_i$ and $r_j$ are on
neighboring lattice points, and $0$ otherwise. Interactions
between neighbors along the chain are not included as their total
only provides a reference point for the other energies. The
interactions between monomer species are tabulated in a matrix
$\Mat{B}$ having mean interaction $B = 0$, and standard deviation
$\delta B = 1$. These values are weighted according to the
fraction of monomers of each species in the system, i.e. $B =
\sum_{k,l} p_k B_{k l} p_l$, and $\delta B^2 = \sum_{k,l} p_k
(B_{k l}- B)^2 p_l$, with $k$ and $l$ taking on the monomer
species types $\alpha$ and $\beta$. With these constraints,
homopolymer effects are removed and the freezing temperature
of the system should be of the order of $\delta B = 1$. We 
first focus on Ising-type interactions, in which $B_{\alpha\alpha} = 
B_{\beta\beta} = 1$ and $B_{\alpha\beta} = 
B_{\beta\alpha} = -1$.

The restriction to compact conformations is partly dictated by
computational constraints, and allows us to fully enumerate much
larger values of $N$ than would be possible otherwise. This choice
is also physically justified since, according to the molten
globule model of  freezing \cite{RMP}, the available states of
proteins at the freezing transition are mostly compact.
Furthermore, all such compact configurations have the
same number of contacts, and therefore energy differences between
configurations are only due to heteropolymeric contributions.
We have selected a compact state with interactions switched off ($B=0$)
as our reference zero energy state (i.e. a non-interacting
compact homopolymer). With this choice, the fluctuation in free energy
over sequences is the heteropolymeric quantity important to sequence
design.

We enumerated chains of length 18 $(3\times 3\times 2)$, 27
$(3\times3\times3)$, and 36 $(3\times3\times4)$.
The ratios  $\alpha:\beta$ in these chains were $9:9$, $14:13,$ and
$18:18$, respectively. We restricted our study to a set 
of 500 sequences each for 18mers and 27mers, and 120 sequences 
for 36mers for reasons of computational
tractability. The enumeration algorithm followed the procedure of Pande
et al \cite{Vijay_Enumeration}. Computations were carried out on two
pentium-II computers and on a cluster at the University of Minnesota
Supercomputing Institute over a period of a few months. 

\begin{figure}
\begin{center}
\epsfig{file=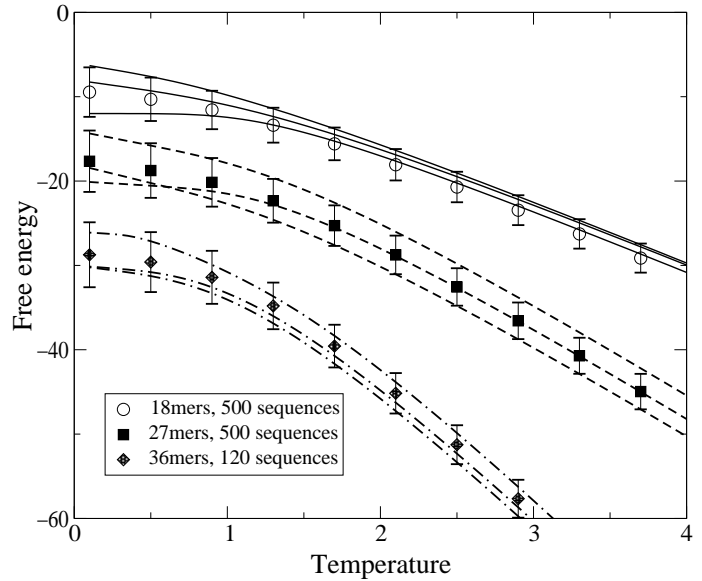, width=3.2in,angle=270}
\end{center}
\caption{ Sequence dependent free energies versus temperature for
three different lengths $N$. The symbols and errorbars indicate
the averaged free energy, and its standard deviation. The solid
lines are the free energy curves for a few sample sequences.
Variations in the free energy are small compared with its absolute
value at all temperatures.}
\end{figure}

Given the limitations of a lattice simulation, we cannot address
questions that depend on the microscopic details of real proteins,
and instead focused on general trends which should be  robust
across different polymer models. The basic test of self-averaging
is whether sequence-dependent variations in  thermodynamic
quantities are significant. Let us first review the general
features of the free energy and its sequence dependent
fluctuations: For any sequence, the free energy $F=E-TS$, is
expected to be linear in temperature at both high and low
temperatures. At high temperatures, all states are accessible, and
the free energy is dominated by $TS_{all}(N)$, where $S_{all}(N)$
is the logarithm of the number of compact conformations of length
$N$. Below its freezing temperature, the free energy is controlled
by the lowest energy states, with a much smaller (possibly zero)
slope of temperature dependence given by the degeneracy of these
states. At these low temperatures, the entropy component of the
free energy is expected to depend strongly on the sequence
\cite{RMP}, though this contribution is small compared to the
energy component, which should be proportional to $N$, and equal
to within $\sqrt{N}$ for all sequences \cite{RMP}. Fig.~(1) shows
a few sample sequences that illustrate this behavior. The error
bars indicate the standard deviation of the free energy $\delta F$
at each temperature, calculated over the ensemble of sequences.
One can see that the behavior is as expected: at high temperature
the curves are parallel and linear in $T$; at low temperature the 
sequence dependence is more important -- in particular below the
freezing temperature, where the slopes of the curves change around
$T \approx 1$.

More importantly, Fig.~(1) shows that the variations in the free 
energy across sequences are significantly less than the absolute 
value of the free energy. In other words, the sequence dependent 
fluctuations are weak and Eq.~(1) is a good approximation. At higher 
temperatures, the relative fluctuations become even less significant, 
because of the greater importance of the $TS$ contribution.

\begin{figure}
\begin{center}
\epsfig{file=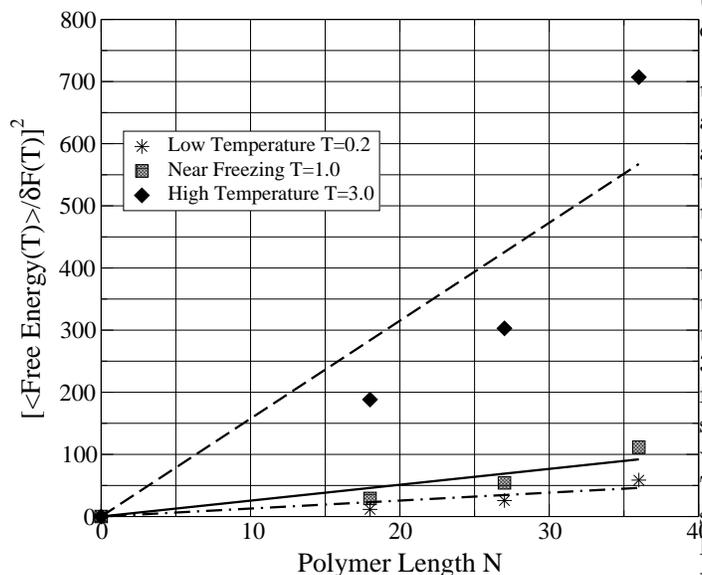, width=3.2in,angle=270}
\end{center}
\caption{The quench-averaged square of the free energy divided by
the free energy variance, as a function of polymer length, at 3
different temperatures. The large values of this quantity, as well
as its increasing trend with $N$, imply that the free energy
variations between sequences will be insignificant for polymers of
hundreds of monomers, the length scale of proteins.}
\end{figure}

In Fig.~(2) we test self-averaging trends by
considering the size dependence of the relative variations in the
free energy. At this stage we should clarify what we
mean by self-averaging, i.e. the conditions that justify the
heteropolymer theories, as well as the fast methods of sequence
design. Heteropolymer theories rely on Eq.~(1), i.e. that
fluctuations are small. More precisely, the standard deviation
among sequences $\delta F(T)$ must be much less than the
average value $\left\langle F(T)\right\rangle$. Design algorithms,
on the other hand, are used to find sequences for which $E$ and
$F$ are of the same magnitude. Therefore, the fast design method
of minimizing just $E$ is a sufficient procedure so long as
sequence-dependent fluctuations of $F$ are much smaller than $F$
itself, and hence much smaller than the sequence dependencies
selected into $E$. Thus the fast design methods will be justified
under the same condition of $\left\langle F\right\rangle/\delta F
\gg 1$. Another trend that we can look for is whether
$\left\langle F\right\rangle/\delta F$ is increasing with $N$. If
this is true, then proteins, which have values of $N$ about an
order of magnitude larger than what we test, should have even
better self-averaging than our lattice models. We indeed expect
such a trend as larger values of $N$ should include more
independent subregions, although this notion is imprecise and
there should be finite size effects \cite{Bastolla}.

As Fig.~(2) shows, the results strongly support self-averaging:
$\left\langle F\right\rangle/\delta F \gg 1$ for all the data points
at all temperatures and polymer lengths.
Furthermore, $\left\langle F\right\rangle/\delta F$ is increasing in
$N$,
which shows that the self-averaging
is even better justified for larger protein-sized polymers.
(The reason for plotting in the square of
$\left\langle F\right\rangle/\delta F$ has to do with extensivity,
as discussed below.)

Figure~(2) is the main result of this paper. It shows that even
for chains as short as $18$ monomers, self-averaging holds for
binary sequences. As real proteins are of the order of $100$ amino
acid units, it seems
likely that self-averaging will be valid for them as well.
Although we are only showing data for Ising interactions, we
believe these results to be valid for all $2\times2$ interaction
matrices with zero mean and unit variance. We have tested several
other matrices, using the simple parameterization suggested in
\cite{Rose_Interaction,Banavar_classes}, with $\theta = 0$,
$\pi/8$, $\pi/4,$ and $3\pi/8$. All of these matrices show similar
trends, and in fact have even higher values for $\left\langle
F\right\rangle/\delta F$ than those presented here. This strongly
suggests that self-averaging is valid independent of the choice of
the interaction matrix. These results should generalize to chains
with more possible monomers (e.g. proteins), so long as the
chains are long enough that the mean and variance of the
monomer-monomer interactions accurately describe the chain
energies. That is, the number of contacts in a conformation should be
at least of the order of the number of possible monomer-monomer
interactions. This would be true, if, as is commonly accepted,
only a few different interactions (e.g. hydrophobicity) are
significant -- though the actual number has been the subject of
some scrutiny \cite{Rose_Interaction}.

A secondary issue related to self-averaging is extensivity.
Self-averaging is traditionally derived from the idea that
sub-regions of the system behave independently\cite{Gredeskul}.
The free energy of the complete system is approximately equal to
the sum of the free energies for many small sub-regions. Each
subregion contains a random realization of local disorder, and if
the system is large, the sum of the free energy over subregions
will be independent of the overall quenched disorder. A
consequence of such independence of subregions is that both the
free energy, and its variance, will be extensive, i.e. linear in
$N$ as $N\to\infty$.

Figure~(3) tests the extensivity of the free energy
at three different temperatures; below, close to, and above,
the freezing transition.
As a rough guide, we have included linear fits that pass
through the four points at $N=0$, $18$, $27$, and $36$.
Note that while the free energy is zero at $N=0$
(a polymer of length zero has no energy or entropy),
the asymptotic linear limit for large $N$ does not have to pass
through this $N=0$ point because of subleading surface terms. 
In order that the different temperatures may be better compared,
the free energies have been divided by temperature,
and compared with their infinite temperature value of $S_{all}(N)$.
The results at $T=3$ are practically indistinguishable from
$S_{all}(N)$, and in fact the dependence on $N$ shows similar trends 
at all three temperatures.
It was shown by Pande et al.\cite{Vijay_Enumeration} that that 
$S_{all}(N)$ has a good linear form when $N$ is extended to lengths
as short as $N=48$. Because of this, we expect extensivity to improve
when $N$ is marginally larger than what we have tested here.

\begin{figure}
\begin{center}
\epsfig{file=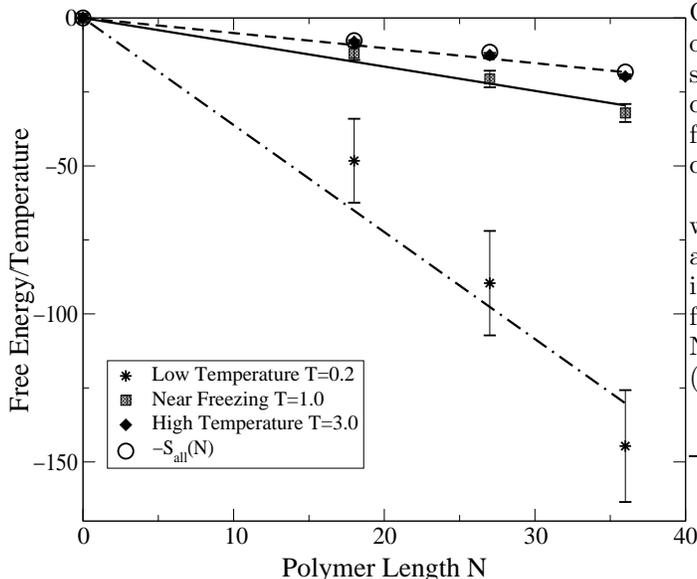, width=3.2in,angle=270}
\end{center}
\caption{The dependence of the free energy on $N$, below, near,
and above the freezing temperature. Each curve has been divided by
the corresponding temperature, so as to compare with the infinite
temperature limit provided by the (logarithm of) the number of
configurations. The deviations from linearity indicate the
importance of finite size effects for protein sized
heteropolymers.}
\end{figure}

Since for a globule made up of independently contributing
subunits we would expect
the variance in the free energy to scale as $N$ as well, the quantity
$\left\langle F(T)\right\rangle^2/\delta F(T)^2$
should be proportional to $N$.
Figure~(2) suggests that this may be the case at least at
low temperatures.
The available data, however, have an upward curvature, and the values
at $N=36$ are systematically higher than our attempted linear fits.
It may well be that, as in the case of entropy calculation in
Ref.\cite{Vijay_Enumeration}, the results for
$\left\langle F(T)\right\rangle^2/\delta F(T)^2$
become linear at marginally higher values of $N$.
Despite these deviations from the expected asymptotic extensivity,
the large magnitude of the plotted values justify self-averaging
according to Eq.~(1).
It is indeed the very deviations from the asymptotic behavior at
these smaller sizes that necessitated the current study, as
it indicates that protein sized objects are not quite extensive
in the thermodynamic sense.

The main conclusion of this work is that sequence-dependent 
fluctuations in the free energy of random heteropolymers are small, 
even at values of $N$ as low as $N = 18$. Qualitatively speaking, 
this means that all random sequences have nearly the same free energy.
There are also indications that the fluctuations decrease in importance
as $N$ increases. These facts together imply that
self-averaging will be a good approximation for protein-sized
heteropolymers.
Although there are deviations from thermodynamic extensivity at this
length scale, the key property of self-averaging is verified.
This latter property is important to sequence design algorithms.
Our results show that sequence design can be carried out without
having to calculate the energy of each tested sequence in all
conformations. Instead, one need only calculate the energy
of each sequence in the desired conformation.
This shortcut vastly reduces the necessary computation time.

We would like to acknowledge illuminating discussions
with the late Professor Toyoichi Tanaka about this work, as well
as inspiration and encouragement from him during the years of
collaboration.  JC acknowledges support from an NSF graduate
fellowship.  MK is supported by NSF grants DMR-98-05833 (at MIT)
and PHY99-07949 (at ITP).

\end{document}